\pdfoutput=1
\documentclass{article}
\usepackage{sage_arxiv,times}

\usepackage{amsmath,amsfonts,bm}

\def\eqref#1{equation~\ref{#1}}

\def\1{\bm{1}}

\newcommand{\test}{\mathcal{D_{\mathrm{test}}}}

\DeclareMathAlphabet{\mathsfit}{\encodingdefault}{\sfdefault}{m}{sl}
\SetMathAlphabet{\mathsfit}{bold}{\encodingdefault}{\sfdefault}{bx}{n}

\usepackage[normalem]{ulem}
\usepackage{amsmath}
\usepackage{amssymb}
\usepackage{booktabs}
\usepackage{multirow}
\usepackage{graphicx}
\usepackage{wrapfig}
\usepackage{capt-of}%
\usepackage[breaklinks=true]{hyperref}
\usepackage{url}
\usepackage{cleveref}
\usepackage{subcaption}
\usepackage{float}
\usepackage{anyfontsize}%
\usepackage{eso-pic}
\definecolor{sagelink}{HTML}{1A4F8A}
\hypersetup{colorlinks=true, urlcolor=sagelink, linkcolor=sagelink, citecolor=sagelink}
\hypersetup{pdftitle={SAGE: Semantic Audio Generative Encoder},
  pdfauthor={Francesco Brigante, Luca Cerovaz, Davide Marincione, Giorgio Strano, Luca Zhou, Emanuele Rodol\`a, Michele Mancusi}}

\title{SAGE: Semantic Audio Generative Encoder}

\author{%
Francesco Brigante\textsuperscript{1} \quad Luca Cerovaz\textsuperscript{1,3} \quad Davide Marincione\textsuperscript{1} \\
{\bfseries Giorgio Strano\textsuperscript{1} \quad Luca Zhou\textsuperscript{1} \quad Emanuele Rodol\`a\textsuperscript{1,3,\dag} \quad Michele Mancusi\textsuperscript{1,2,\dag}} \\[0.55em]
{\normalfont\small\textsuperscript{1}Sapienza University of Rome, Italy} \quad \textsuperscript{2}Moises Systems, Inc.\quad\textsuperscript{3}Paradigma}

\begin{document}

\AddToShipoutPictureFG*{\AtPageUpperLeft{\put(\LenToUnit{\dimexpr\paperwidth-1in\relax},\LenToUnit{-0.85in}){\includegraphics[width=0.6in]{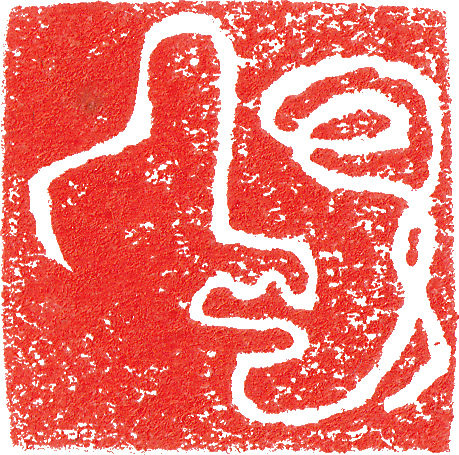}}}}

\maketitle

{\let\thefootnote\relax\footnotetext{Correspondence to: \texttt{brigante@di.uniroma1.it}}}
{\let\thefootnote\relax\footnotetext{\textsuperscript{\dag}Equally advising}}

\begin{abstract}
Audio autoencoders compress waveforms into compact latent representations that serve as the interface between raw audio and downstream models. Current systems navigate a three-way trade-off between reconstruction quality, semantic structure of the latent space, and inference speed, typically favoring one or two of these at the expense of the others. This paper introduces SAGE, \emph{Semantic Audio Generative Encoder}: a compact variational autoencoder, trained solely on publicly available music, that shapes its latent by distilling embeddings from a pretrained audio-text model. This $105$M-parameter model runs about $3\times$ faster than Stable Audio Open and reaches the listening-test quality of SAME-L, an autoencoder $8\times$ larger and over $2\times$ slower, while surpassing both on objective perceptual and distributional metrics of reconstruction. Furthermore, it sets the state of the art on all nineteen probing tasks of latent semantics, in domain and out of domain. These results establish SAGE as a lightweight audio autoencoder that strikes the best balance of the three-way trade-off among those we evaluate, combining high reconstruction fidelity, state-of-the-art semantic structure, and fast inference.

\end{abstract}

\section{Introduction}
\label{sec:intro}

\begin{wrapfigure}[17]{r}{0.46\linewidth}
  \centering
  \vspace{-1.2\baselineskip}
  \includegraphics[width=\linewidth]{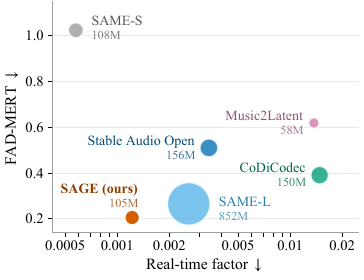}
  \caption{Distributional fidelity against inference cost on the MoisesDB mixtures; marker area is the parameter count.}
  \label{fig:efficiency}
\end{wrapfigure}

State-of-the-art music generation follows the latent diffusion recipe \citep{ldm}. An autoencoder compresses the waveform into a compact latent. A generative model then learns to sample in that latent space. The autoencoder therefore runs at both ends of the pipeline since it encodes the training corpus of the generator and decodes every generated sample. While few-step samplers \citep{consistencymodels} cut sampling from tens or hundreds of denoising steps to only a few, the decoder still runs at full cost on every sample and accounts for a large share of inference latency. For interactive, editing, and long-form applications, its cost is thus critical.

Yet current music autoencoders force a choice between speed and fidelity. Stable Audio Open~\citep{stableaudioopen}, a convolutional stack on the waveform at the sample rate, trails in fidelity. SAME~\citep{same} reaches the best fidelity of the systems we compare and is the only one that shapes its latent semantically, but its large configuration has $852$M parameters, and the small one, distilled from it to run faster, pays for the speed in fidelity. Music2Latent~\citep{music2latent} and CoDiCodec~\citep{codicodec} operate on the spectrogram and decode through a consistency model; like Stable Audio Open, they optimize the latent for reconstruction alone, so any semantic structure in it is incidental. No system offers speed, the quality of SAME-L and a semantically organized latent at once.

A spectrogram is a two-dimensional time-frequency signal. Vision already offers efficient hierarchical encoders for such signals. Swin \citep{swinv1,swinv2} restricts attention to local windows, so its cost grows linearly with input size. Its stages downsample progressively and build multiscale features. In audio, this backbone is established for classification \citep{htsat}, not for generative autoencoding. Latent semantics has a similar gap: in images, aligning generative representations with pretrained encoders speeds up training and improves samples \citep{yu2025representationalignmentgenerationtraining,vavae}. In audio, the same idea has relied on large models or closed codecs, costing some fidelity.

\begin{figure}[!ht]
  \centering
  \includegraphics[width=\linewidth]{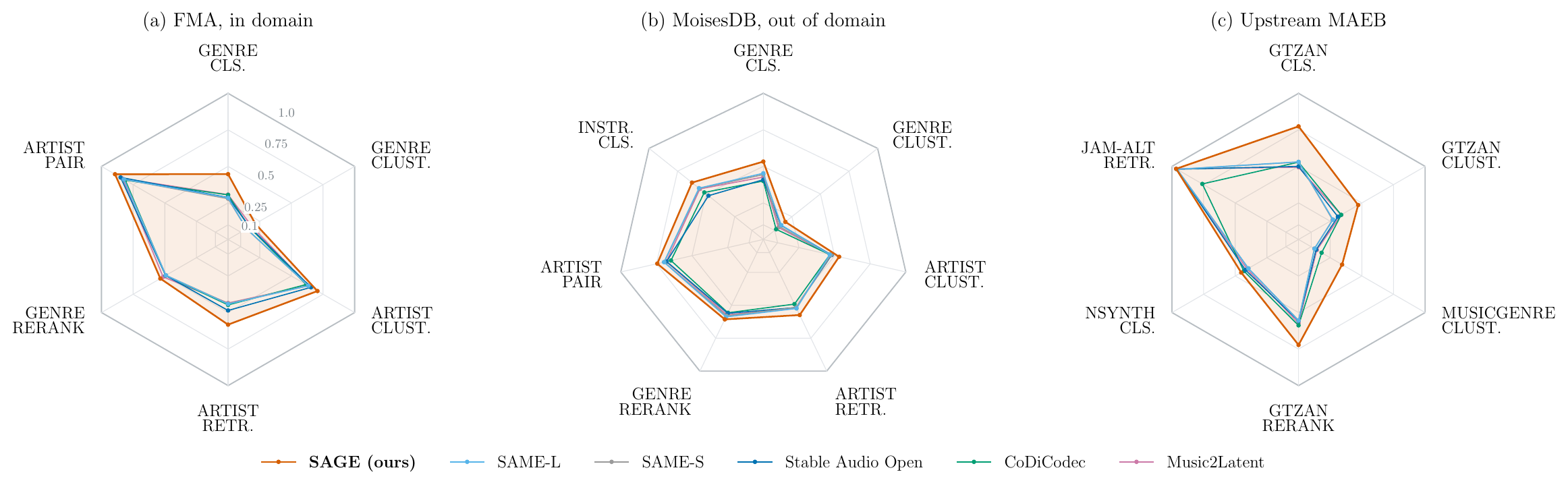}
  \caption{Probing on the nineteen tasks, one panel per block; higher is better on every task.}
  \label{fig:maeb-radar}
\end{figure}

We propose SAGE, a compact VAE that encodes and compresses the complex STFT of stereo music with a hierarchical SwinV2 backbone \citep{swinv2} and distills a frozen CLAP encoder \citep{clap} into its latent, yielding a semantically organized representation that directly predicts the complex spectrum without the need for vocoders or phase reconstruction.
We evaluate SAGE along two axes: reconstruction, through perceptual and distributional metrics, and latent semantics, through nineteen music probing tasks in the format of MAEB~\citep{maeb}, six from the benchmark itself and thirteen we build on FMA~\citep{fma} and MoisesDB~\citep{moisesdb}. Our 105M-parameter model sits at the corner of the trade-off in \Cref{fig:efficiency}: it runs about three times faster than Stable Audio Open and is outrun only by the distilled SAME-S, whose Fr\'echet distance is several times higher, and it matches the subjective quality of SAME-L, the strongest baseline, at an eighth of its parameters and less than half its inference time. On objective perceptual and distributional metrics it surpasses both, in domain and out of domain, and it leads every one of the nineteen probing tasks (\Cref{fig:maeb-radar}).

Our contributions are:
\begin{itemize}
    \item \textbf{A spectrogram-native hierarchical autoencoder.} We adapt a
    SwinV2 encoder-decoder from images to the complex STFT, with rectangular
    patches and attention windows. At $105$M parameters, the result runs about three times faster than Stable Audio Open and matches SAME-L in a listening test, surpassing both on perceptual and distributional reconstruction metrics.
    
    \item \textbf{Semantic distillation on a delayed schedule.} We distill CLAP into the latent, delayed during training so that reconstruction is established before alignment begins; we sweep the introduction of the latent term, the loss weight and the teacher, and identify the knee at which semantic structure is gained at the least cost in fidelity.

    \item \textbf{A thorough evaluation of audio autoencoders.} We measure every
    compared system through a single harness on five corpora. Reconstruction is
    scored with perceptual, distributional and sample-exact metrics, complemented by
    a MUSHRA listening test \citep{mushra} on held-out commercial music, in which SAGE and SAME-L are statistically indistinguishable; latent semantics is scored
    with nineteen probing tasks in the MAEB format, thirteen of them instantiated on the
    same corpora. 
\end{itemize}

Weights, code and the evaluation harness are available at
\url{https://github.com/francescobrigante/SAGE}, with the project page at
\url{https://sage-music.pages.dev/}.

\section{Method}
\label{sec:method}

\subsection{Audio-Adapted Hierarchical Autoencoder}
\label{sec:backbone}

SAGE is a hierarchical VAE~\citep{vae} that compresses an audio signal $\mathbf{X}$ into a latent $x$ by progressively shrinking the input via SwinV2 \citep{swinv2} blocks. Its input is the stacked $\mathbf{X} \in \mathbb{R}^{4 \times F \times T}$ representation of the left and right channels, and real and imaginary parts, of a dual-complex spectrogram,  obtained via a short-time Fourier transform (STFT) with a $2048$-sample Hann window and a hop of $512$ samples, of a $44.1$\, kHz stereo waveform. To ease the blocks' compression, we drop the Nyquist bin and rescale the spectrum's magnitudes with a power law, as in CoDiCodec. We predict the complex spectrum directly, so no vocoder or phase reconstruction is needed.

\begin{figure}[!h]
  \centering
  \includegraphics[width=\linewidth]{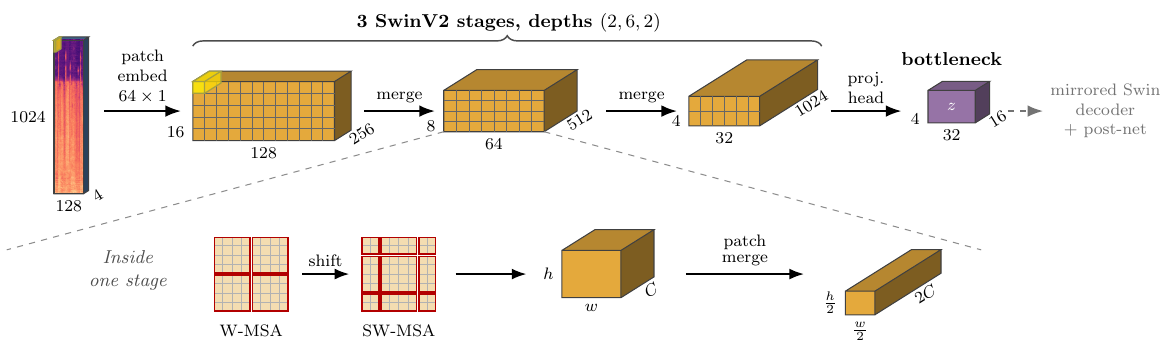}
  \caption{The SAGE encoder: (top) the tensor pipeline through the three SwinV2 stages to the latent bottleneck; (bottom) the internals of one stage.}
  \label{fig:architecture}
\end{figure}

Due to the nature of audio spectrograms, we introduce three adaptations to the SwinV2 architecture (\Cref{fig:architecture}): strongly rectangular patches, matched by rectangular attention windows, and a shallow three-stage hierarchy. The activation function in the MLP is replaced by a SwiGLU \citep{swiglu}, and attention is made exclusive \citep{xsa}, orthogonalizing each token's attention output against its own value vector, so that it gathers only information complementary to the token itself. Otherwise, blocks retain their original design (scaled-cosine attention, log-spaced continuous position bias, post-residual normalization).

The encoder output $x$ defines a diagonal Gaussian posterior, and the VAE bottleneck samples the latent $z$ from it. A Kullback-Leibler term $\mathcal{L}_{\mathrm{KL}}$, averaged over the batch and the latent grid, pulls this posterior towards a standard normal prior. As in other autoencoders that feed into a second-stage generative model, the weight on this term is deliberately small (\Cref{sec:curriculum}): in the rate-distortion trade-off it governs \citep{betavae}, a light penalty confines the latent to a smooth, approximately standardized region without discarding the fine spectral structure the decoder must recover.
Throughout our experiments, we hold the latent at the compression ratio of Stable Audio Open, so the two are compared at the same $\times 64$ budget.
Finally, a mirrored decoder reverses the encoder's hierarchy and produces the reconstruction $\hat{\mathbf{X}} = D(z)$, followed by a lightweight residual refinement layer that reduces discontinuities between reconstructed frequency bands.

\subsection{Semantic Latent Distillation}
\label{sec:distillation}
Reconstruction quality, the distortion axis of the rate-distortion-perception trade-off, says nothing about how the latent is organized, yet that organization is what probing, retrieval and classification consume directly and what a downstream generative model must learn from (\Cref{sec:semantic-alignment}).
SAGE therefore distills \citep{distillation} its latent towards the embedding space of CLAP, a contrastive language-audio model whose audio arm, HTS-AT \citep{htsat}, maps a sample to a single $512$-dimensional embedding aligned with natural-language descriptions of its content. During training, CLAP remains frozen, and the alignment follows the clip-level scheme of SALAD-VAE~\citep{salad}; the resulting branch is the upper pathway of \Cref{fig:curriculum}a.

Folding $z \in \mathbb{R}^{16 \times 4 \times 32}$ along frequency gives $16 \cdot 4 = 64$ channels over $32$ frames, averaged over time into a clip descriptor $\bar{z} \in \mathbb{R}^{64}$. A learned linear head $\phi$ projects the descriptor into the teacher's space, and the loss maximizes its cosine similarity to the teacher embedding $e$ of the same clip
\begin{equation}
  \mathcal{L}_{\mathrm{sem}} \;=\; 1 - \cos\!\big(\phi(\bar{z}),\, e\big),
  \label{eq:distill}
\end{equation}
bounded in $[0, 2]$, which keeps its scale commensurate with the other terms. The teacher consumes the same segment downmixed to mono and resampled to its native $48$\,kHz. Teacher and head are used only during training and discarded at inference; furthermore, the head is optimized at a lower learning rate than the rest of the model, since a head that adapts too quickly would shoulder most of the alignment, discouraging the encoder from pursuing it and making it unaligned at inference time.

\paragraph{Detached warm-up.} For the first $s_0$ training steps the descriptor entering \eqref{eq:distill} is $\mathrm{sg}[\bar{z}]$, the stop-gradient of $\bar{z}$: the latent develops its structure under the reconstruction and KL objectives alone, while $\phi$, active from the first step, learns to track it. From step $s_0$ on, $\bar{z}$ enters undetached and $\mathcal{L}_{\mathrm{sem}}$ reshapes a representation that already encodes the signal. Later placement of the gate $s_0$ favors reconstruction and earlier placement favors semantic structure; we set it at the empirical sweet spot of Section~\ref{sec:ablations}.

\subsection{Two-Phase Training Curriculum}
\label{sec:curriculum}

\begin{figure}[t]
    \centering
    \subcaptionbox[Short Subcaption]{%
        Pretraining%
        \label{subfig:curriculum_phase1}%
    }
    [%
        0.418\textwidth%
    ]%
    {%
        \includegraphics[width=0.418\textwidth]%
        {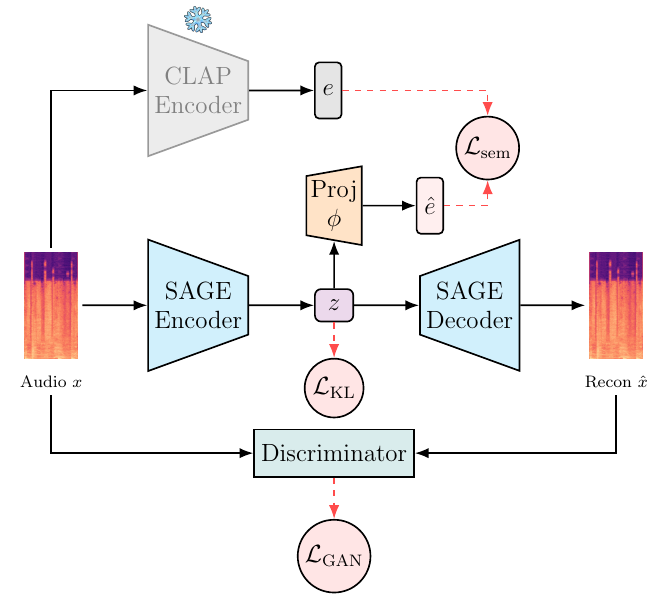}%
    }%
    \hfill
    \subcaptionbox[Short Subcaption]{%
        Decoder fine-tuning%
        \label{subfig:curriculum_phase2}%
    }
    [%
        0.5035\textwidth%
    ]%
    {%
        \includegraphics[width=0.5035\textwidth]%
        {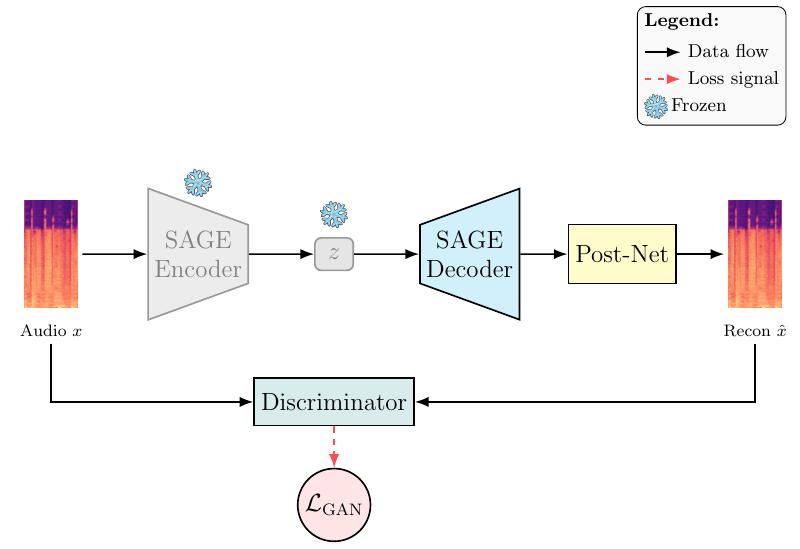}%
    }%
    \caption[Short Caption]{The two training phases of SAGE.}
    \label{fig:curriculum}
\end{figure}

A decoder trained only on average spectral distances regresses to the mean of the reconstructions compatible with a latent, which is smoother than any of them and lacks the high-frequency detail that carries much of the perceived realism. Following neural audio codecs, we add an adversarial objective \citep{gan} against the composite discriminator of WavTokenizer~\citep{wavtokenizer}, applied to the waveform obtained by inverting the predicted spectrogram. Since this discriminator judges each channel independently, we also feed it the mid $M = L + R$ and side $S = L - R$ signals. We use the relativistic RpGAN formulation~\citep{rgan,rpgan}, which rates real audio against its own reconstruction; the generator receives the adversarial term $\mathcal{L}_{\mathrm{adv}}$ and a feature-matching term $\mathcal{L}_{\mathrm{fm}}$~\citep{melgan}. Fidelity is enforced by three terms (Appendix~\ref{app:losses}): $\mathcal{L}_{\mathrm{STFT}}$, a squared error on the power-compressed complex spectrogram, penalizing magnitude and phase jointly; $\mathcal{L}_{\mathrm{mel}}$~\citep{parallelwavegan}, a multi-resolution mel distance on the reconstructed waveform; and $\mathcal{L}_{\mathrm{SD}}$, the sum-and-difference multi-resolution STFT loss of Stable Audio Open. The last term is needed for the stereo image: a squared error weighs the low-energy side only by its share of the energy, and without $\mathcal{L}_{\mathrm{SD}}$ the side is reconstructed about $12$\,dB too quiet (Appendix~\ref{app:stereo}).

\paragraph{Phase 1: pretraining.} The full system of \Cref{subfig:curriculum_phase1} trains from scratch on
\begin{equation}
  \mathcal{L} = \lambda_{\mathrm{STFT}}\mathcal{L}_{\mathrm{STFT}} + \lambda_{\mathrm{mel}}\mathcal{L}_{\mathrm{mel}} + \lambda_{\mathrm{SD}}\mathcal{L}_{\mathrm{SD}} + \lambda_{\mathrm{KL}}\mathcal{L}_{\mathrm{KL}} + \lambda_{\mathrm{sem}}\mathcal{L}_{\mathrm{sem}} + \lambda_{\mathrm{adv}}\mathcal{L}_{\mathrm{adv}} + \lambda_{\mathrm{fm}}\mathcal{L}_{\mathrm{fm}},
  \label{eq:total}
\end{equation}
with $\mathcal{L}_{\mathrm{sem}}$ under the detached warm-up of Section~\ref{sec:distillation}. Reconstruction dominates and the adversarial pair is kept well below it (loss weights in Appendix~\ref{app:hparams}, Table~\ref{tab:hparams}). The discriminator is active from the first step, which suits the WavTokenizer discriminator (Appendix~\ref{app:ablations}). An exponential moving average (EMA) of the weights is used for evaluation and to initialize the next phase.

\paragraph{Phase 2: decoder fine-tuning.} We reload the EMA weights and freeze the encoder (\Cref{subfig:curriculum_phase2}), so the latent consumed by downstream models stays exactly as pretraining left it. The KL and semantic terms are switched off (Appendix~\ref{app:hparams}), a zero-initialized post-net is attached, and the discriminator is re-initialized. The decoder alone then trains further, gaining perceptual quality.

\section{Experiments}
\label{sec:experiments}

\subsection{Experimental Setup}
\label{sec:setup}

\paragraph{Model configuration.} SAGE's architecture (\Cref{sec:backbone}) is instantiated with an embedding size of $C = 256$ in the first block, three stages of depths $(2, 6, 2)$, $(8, 16, 32)$ attention heads, $64 \times 1$ patches, $4 \times 32$ attention windows, $C_z = 16$ latent channels and stochastic depth at rate $0.1$ in encoder and decoder, the decoder carrying in addition the residual post-net, composed of two $64$-channel convolutions. The total size, barring the CLAP teacher, the discriminator and the projection head $\phi$, is $104.6$M parameters. The teacher is the LAION-CLAP checkpoint trained on music and general audio.

\paragraph{Data.} SAGE is trained on $\approx$10.5K hours of 44.1\,kHz stereo audio from three public corpora: the FMA-full training split, the MTG-Jamendo training partition \citep{jamendo} (excluding tracks used by the Song Describer Dataset), and M4Singer \citep{m4singer}, whose solo-singing recordings expose the model to isolated voice alongside full mixes. To avoid over-sampling M4Singer's many short recordings, each epoch uses a different quarter of it alongside all of FMA and Jamendo ($\approx$122.4K tracks, one random 1.5\,s crop each).

Reconstruction is evaluated on five held-out sets: the FMA test split (11{,}263 clips of 30\,s, $20\times$ the training window) as the in-domain benchmark; MoisesDB mixtures and isolated stems (1{,}998 and 1{,}546 windows of 10\,s), professionally produced multitracks that are out of domain; and MusicCaps \citep{musiccaps} (964 clips of 10\,s) and the Song Describer Dataset \citep{songdescriber} (8{,}364 chunks of 10\,s), the benchmarks of Music2Latent/CoDiCodec and Stable Audio Open/SAME, respectively, so that every baseline is also scored on its authors' own choice of corpus. Semantic probing uses the FMA test split, 1{,}611 chunks of 30\,s from MoisesDB mixtures and stems (the stems enable source-level tasks), and the external corpora of \Cref{sec:protocol}.
\paragraph{Training.} Phase~1 trains from scratch for $500$ epochs ($\approx 239$k generator updates) on $16$ A100 GPUs at a global batch of $128$ segments, with AdamW~\citep{adamw} at a peak learning rate of $10^{-3}$ under an inverse square-root schedule, with TF32 matrix multiplications. The discriminator and the projection head have their own optimizers; the head runs at the slow constant rate required by \Cref{sec:distillation}. Phase~2 keeps hardware, batch and corpus fixed, lowers the generator learning rate to $10^{-4}$ with a restarted schedule, and runs for $992$ epochs. Appendix~\ref{app:hparams} gives the full optimization settings. The two phases cost $1{,}536$ and $3{,}043$ GPU-hours, respectively.

\paragraph{Baselines.} In \Cref{tab:specs} we compare against five open-weight continuous-latent audio autoencoders; Stable Audio Open is the convolutional waveform-domain VAE whose compression budget SAGE adopts. SAME is released in two configurations and we evaluate both: SAME-L, the flagship, is by far the largest system in the comparison, and SAME-S is a $108$M-parameter variant distilled from it to run on CPU, which puts it at SAGE's size. Both share the $256$-dimensional latent and the semantic shaping of SAME (\Cref{sec:semantic-alignment}), making them the only other systems that organize their latent on purpose. CoDiCodec, run in its continuous mode, and Music2Latent are consistency autoencoders on the complex STFT, the former at twice the compression of Stable Audio Open and SAGE; Music2Latent treats the two channels independently rather than modeling them jointly, and is run on both, so its ratio holds per channel. Every baseline is run from its public weights through the same harness, one call per clip with no external chunking or overlap-add. Timings use each authors' implementation at batch size one, compiled, in TF32 for every model; Appendix~\ref{app:timing} gives the protocol, eager timings and other precisions. SALAD-VAE, whose objective \Cref{sec:distillation} adapts, is not a baseline: it has released neither weights nor code, and targets general audio.

\begin{table}[!ht]
  \centering
  \small
  \caption{Compiled TF32 time per clip on one RTX PRO 6000 Blackwell (Appendix~\ref{app:timing}); best in bold, second underlined. Data hours as reported; compression is waveform samples per latent scalar.}
  \label{tab:specs}
  \setlength{\tabcolsep}{2.5pt}
  \begin{tabular}{@{}lrrlrrrrrr@{}}
    \toprule
    & & & & & & \multicolumn{2}{c}{MoisesDB, $10$\,s} & \multicolumn{2}{c}{FMA, $30$\,s} \\
    \cmidrule(lr){7-8} \cmidrule(lr){9-10}
    Model & Params & Data (h) & Domain & Latent & Comp. & ms $\downarrow$ & vs SAGE & ms $\downarrow$ & vs SAGE \\
    \midrule
    \textbf{SAGE (ours)} & $105$M & $10{,}460$ & STFT & $64$ & $\times 64$ & $\underline{12.2}$ & $\times1.00$ & $\underline{30.6}$ & $\times1.00$ \\
    \midrule
    SAME-L & $852$M & $\approx 19{,}500$ & waveform & $256$ & $\times 32$ & $25.9$ & $\times2.12$ & $78.3$ & $\times2.56$ \\
    SAME-S & $108$M & $\approx 19{,}500$ & waveform & $256$ & $\times 32$ & $\mathbf{5.77}$ & $\times0.47$ & $\mathbf{14.7}$ & $\times0.48$ \\
    Stable Audio Open & $156$M & $\approx 7{,}300$ & waveform & $64$ & $\times 64$ & $33.9$ & $\times2.78$ & $100.1$ & $\times3.28$ \\
    CoDiCodec & $\approx 150$M & $\approx 4{,}030$ & STFT & $64$ & $\times 128$ & $148.2$ & $\times12.1$ & $413.8$ & $\times13.5$ \\
    Music2Latent & $58$M & $\approx 3{,}800$ & STFT & $64$ & $\times 64$ & $137.4$ & $\times11.3$ & $405.3$ & $\times13.3$ \\
    \bottomrule
  \end{tabular}
\end{table}

\subsection{Evaluation Protocol}
\label{sec:protocol}

\paragraph{Reconstruction metrics.} We report three families of reconstruction metrics and give priority to the first two. The perceptual family is the cosine similarity between the CLAP embeddings of a clip and of its reconstruction. We compute it under the music checkpoint, the teacher of \Cref{sec:distillation}, and under the general-audio checkpoint, which plays no part in training. A MUSHRA listening test completes the family with the human judgment these similarities approximate. The distributional family is the Fr\'echet audio distance \citep{fad} between the reference and the reconstruction distributions, computed in three embedding spaces, framewise MERT \citep{mert} features, whole-clip embeddings from that same general-audio checkpoint and PANN \citep{pann} features, since the ranking it induces is known to depend on the space in which it is taken \citep{fadtk}. Together they answer the question we care about, whether a reconstruction sounds plausible and is hard to tell from the original. The third family is sample-exact and collects the signal-to-distortion ratio \citep{bsseval} and the multi-resolution STFT distance that the objective of \Cref{sec:curriculum} minimizes directly. We keep these in every table and read them as diagnostics, since they charge for inaudible phase rotations and favor a model that regresses the waveform sample by sample over one that predicts a spectrum.

\paragraph{Semantic probing.} We freeze the encoder and probe its latents with the nineteen audio-only music tasks. On the corpora of \Cref{sec:setup}, so that both axes are evaluated on the same data, they are genre classification, genre and artist clustering, artist retrieval, genre reranking and artist pair classification on the FMA test split and the MoisesDB mixtures, plus instrument classification on the isolated MoisesDB stems, a source-level task that mixtures cannot pose. The remaining six are the upstream tasks of MAEB on their original datasets, which none of the compared models have seen in training. For every model we fold the latent along frequency into channels and average it over time. Classification tasks fit a track-grouped, cross-validated logistic regression while clustering, retrieval, reranking and pair classification train nothing and read the pooled latent directly.

\subsection{Reconstruction}
\label{sec:reconstruction}

\begin{figure}[H]
  \centering
  \includegraphics[width=0.874\linewidth]{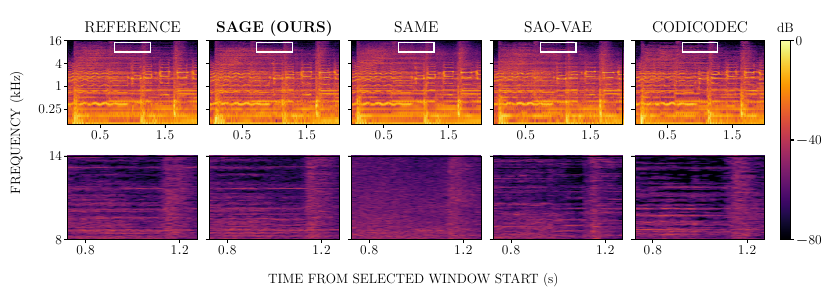}
  \vspace{-3mm}
  \caption{
    STFT power spectrograms of a selected music excerpt.
    Top: reference and reconstructions.
    Bottom: the same highlighted region enlarged for each model.
  }
  \label{fig:reconstruction-detail}
\end{figure}

\begin{table}[!ht]
  \centering
  \small
  \setlength{\tabcolsep}{4pt}
  \caption{Reconstruction on the five evaluation sets, best per column within each set in bold.}
  \label{tab:recon}
  \begin{tabular}{@{}clccccccc@{}}
    \toprule
    & & \multicolumn{3}{c}{FAD $\downarrow$} & \multicolumn{2}{c}{CLAP $\uparrow$} & & \\
    \cmidrule(lr){3-5} \cmidrule(lr){6-7}
    & Model & MERT & CLAP & PANN & music & audio & SDR $\uparrow$ & dSTFT $\downarrow$ \\
    \midrule
    \multirow{6}{*}{\rotatebox[origin=c]{90}{\scriptsize\itshape\shortstack{FMA\\test}}}
    & \textbf{SAGE (ours)} & $\mathbf{0.113}$ & $\mathbf{0.016}$ & $\mathbf{1.422}$ & $\mathbf{0.956}$ & $\mathbf{0.960}$ & $5.28$ & $\mathbf{0.929}$ \\
    & SAME-L                & $0.210$ & $0.038$ & $3.470$ & $0.897$ & $0.927$ & $\mathbf{10.91}$ & $1.746$ \\
    & Stable Audio Open    & $0.319$ & $0.044$ & $2.891$ & $0.892$ & $0.911$ & $5.64$ & $1.342$ \\
    & CoDiCodec            & $0.283$ & $0.026$ & $3.243$ & $0.907$ & $0.934$ & $1.78$ & $1.379$ \\
    & SAME-S               & $0.649$ & $0.082$ & $7.169$ & $0.839$ & $0.870$ & $8.80$ & $1.675$  \\
    & Music2Latent         & $0.396$ & $0.045$ & $4.599$ & $0.862$ & $0.906$ & $1.21$ & $1.531$ \\
    \midrule
    \multirow{6}{*}{\rotatebox[origin=c]{90}{\scriptsize\itshape\shortstack{MoisesDB\\mixtures}}}
    & \textbf{SAGE (ours)} & $\mathbf{0.205}$ & $\mathbf{0.026}$ & $1.438$ & $\mathbf{0.966}$ & $\mathbf{0.971}$ & $4.79$ & $0.857$ \\
    & SAME-L                & $0.265$ & $0.033$ & $\mathbf{1.277}$ & $0.954$ & $0.966$ & $\mathbf{9.49}$ & $\mathbf{0.816}$ \\
    & Stable Audio Open    & $0.510$ & $0.118$ & $2.477$ & $0.933$ & $0.944$ & $5.02$ & $0.956$ \\
    & CoDiCodec            & $0.391$ & $0.130$ & $11.531$ & $0.930$ & $0.923$ & $1.84$ & $0.848$ \\
    & SAME-S               & $1.024$ & $0.094$ & $4.266$ & $0.914$ & $0.930$ & $7.97$ & $0.839$ \\
    & Music2Latent         & $0.619$ & $0.318$ & $3.362$ & $0.930$ & $0.945$ & $1.22$ & $0.943$ \\
    \midrule
    \multirow{6}{*}{\rotatebox[origin=c]{90}{\scriptsize\itshape\shortstack{MoisesDB\\stems}}}
    & \textbf{SAGE (ours)} & $\mathbf{0.135}$ & $\mathbf{0.027}$ & $\mathbf{1.692}$ & $\mathbf{0.962}$ & $\mathbf{0.965}$ & $6.69$ & $\mathbf{1.058}$ \\
    & SAME-L                & $0.559$ & $0.087$ & $4.141$ & $0.888$ & $0.911$ & $\mathbf{12.89}$ & $1.398$ \\
    & Stable Audio Open    & $0.552$ & $0.121$ & $3.198$ & $0.934$ & $0.942$ & $7.05$ & $1.190$ \\
    & CoDiCodec            & $0.611$ & $0.140$ & $4.418$ & $0.882$ & $0.903$ & $4.00$ & $1.295$ \\
    & SAME-S               & $1.348$ & $0.145$ & $7.713$ & $0.852$ & $0.857$ & $9.70$ & $1.357$ \\
    & Music2Latent         & $1.650$ & $0.331$ & $6.564$ & $0.834$ & $0.854$ & $3.52$ & $1.646$ \\
    \midrule
    \multirow{6}{*}{\rotatebox[origin=c]{90}{\scriptsize\itshape MusicCaps}}
    & \textbf{SAGE (ours)} & $0.306$ & $\mathbf{0.029}$ & $\mathbf{3.169}$ & $\mathbf{0.930}$ & $\mathbf{0.957}$ & $5.09$ & $\mathbf{1.229}$ \\
    & SAME-L                & $0.508$ & $0.092$ & $3.529$ & $0.870$ & $0.914$ & $\mathbf{10.56}$ & $1.741$ \\
    & Stable Audio Open    & $0.529$ & $0.125$ & $5.171$ & $0.867$ & $0.909$ & $5.47$ & $1.258$ \\
    & CoDiCodec            & $\mathbf{0.284}$ & $0.056$ & $5.635$ & $0.918$ & $0.940$ & $1.83$ & $1.239$ \\
    & SAME-S               & $0.957$ & $0.134$ & $6.484$ & $0.828$ & $0.873$ & $8.56$ & $1.656$ \\
    & Music2Latent         & $0.582$ & $0.287$ & $5.713$ & $0.887$ & $0.928$ & $1.12$ & $1.406$ \\
    \midrule
    \multirow{6}{*}{\rotatebox[origin=c]{90}{\scriptsize\itshape\shortstack{Song\\Describer}}}
    & \textbf{SAGE (ours)} & $\mathbf{0.098}$ & $\mathbf{0.011}$ & $\mathbf{0.888}$ & $\mathbf{0.969}$ & $\mathbf{0.976}$ & $6.31$ & $\mathbf{0.878}$ \\
    & SAME-L                & $0.186$ & $0.037$ & $1.497$ & $0.940$ & $0.955$ & $\mathbf{12.17}$ & $1.451$ \\
    & Stable Audio Open    & $0.349$ & $0.126$ & $2.417$ & $0.924$ & $0.942$ & $6.86$ & $1.213$ \\
    & CoDiCodec            & $0.290$ & $0.085$ & $8.397$ & $0.922$ & $0.920$ & $2.42$ & $1.178$ \\
    & SAME-S               & $0.648$ & $0.087$ & $4.732$ & $0.896$ & $0.915$ & $9.97$ & $1.399$ \\
    & Music2Latent         & $0.408$ & $0.276$ & $2.847$ & $0.915$ & $0.935$ & $2.24$ & $1.290$ \\
    \bottomrule
  \end{tabular}
\end{table}

\Cref{tab:recon} reports reconstruction on the five evaluation sets of \Cref{sec:setup}. SAGE holds the lowest
Fr\'echet distance and the highest CLAP similarity on twenty-three of the twenty-five measurements
the five sets provide, on twenty-four of them against SAME-L, a model eight times larger, and on every one against SAME-S; the
exceptions are FAD-MERT on MusicCaps, where CoDiCodec is ahead, and FAD-PANN on the MoisesDB
mixtures, where SAME-L is.
\Cref{fig:efficiency} places these results against inference cost. Stable Audio Open runs about three times slower than SAGE and trails it on every perceptual and distributional measurement; CoDiCodec and Music2Latent trail as well while running more than eleven times slower; SAME-L is the only system that listeners rate on par with SAGE (\Cref{tab:mushra}), at eight times the parameters and more than twice the inference time. SAGE thus runs faster than both while matching the perceived quality of SAME-L. \Cref{fig:reconstruction-detail} shows the same trend on a single excerpt.

The only faster system, SAME-S, pays for its speed in the distributional metrics: its FAD-MERT is three to ten times SAGE's, the highest of any system on four of the five sets. Its authors' listening test agrees, rating it below every other system they tested, Stable Audio Open included ($66.1$ vs.\ $73.3$), despite a higher SI-SDR ($9.6$ vs.\ $6.2$\,dB). The Fr\'echet distances in the MERT and PANN spaces place it behind Stable Audio Open as the listeners do, which is why \Cref{sec:protocol} gives them priority and keeps SDR as a diagnostic.

The MUSHRA test uses ten $10$\,s excerpts of commercial music held out of training, comparing SAGE against SAME-L, Stable Audio Open and CoDiCodec, together with a hidden reference and a $3.5$\,kHz low-pass anchor. Music2Latent and SAME-S are omitted to shorten the test: the first trails all four on FAD-MERT on every set of \Cref{tab:recon}, and the second is rated below Stable Audio Open in the test of its own authors. Participants are filtered on their ability to correctly rate the hidden reference. \Cref{tab:mushra} reports the filtered results, for a total of $21$ raters. Two groups emerge. SAGE and SAME-L score $81.6$ and $81.8$ with overlapping 95\% confidence intervals, so the listening test cannot separate them, although SAME-L has eight times the parameters and more than twice the inference time. Stable Audio Open and CoDiCodec form a second group and trail by fifteen to seventeen points; in particular, SAGE scores 17 points above Stable Audio Open while running about three times faster.
\begin{table}[!ht]
  \centering
  \fontsize{8.55}{10.45}\selectfont%
  \caption{Mean MUSHRA score $\pm$ 95\% CI, with parameters / RTF (TF32, compiled, as in \Cref{tab:specs}) under each model; $38$ participants, $21$ after filtering. The hidden reference scored $97.9\pm0.7$ and the $3.5$\,kHz low-pass anchor $15.4\pm2.0$.}
  \label{tab:mushra}
  \begin{tabular}{cccc}
    \toprule
    \begin{tabular}{c}
         \textbf{SAGE (ours)} \\
         $105$M / $0.0012$
    \end{tabular} & \begin{tabular}{c}
         SAME-L \\
         $852$M / $0.0026$
    \end{tabular} & \begin{tabular}{c}
         Stable Audio Open \\
         $156$M / $0.0034$
    \end{tabular} & \begin{tabular}{c}
         CoDiCodec \\
         $150$M / $0.0148$
    \end{tabular} \\
    \midrule
    $\mathbf{81.6\pm2.7}$ & $\mathbf{81.8\pm2.6}$ & $64.6\pm3.7$ & $66.4\pm3.5$ \\
    \bottomrule
  \end{tabular}
\end{table}

\subsection{Semantic Probing}
\label{sec:probing}
\noindent
\begin{minipage}[t]{0.65\linewidth}
  \setlength{\parindent}{0pt}
  \Cref{fig:maeb-radar} reports the nineteen tasks in their three blocks, with the absolute scores and a CLAP-oracle reference row in \Cref{tab:maeb-full} of Appendix~\ref{app:maeb}.
Every score is the metric MAEB defines for its task type (\Cref{tab:maeb-full}); all are bounded by one, so the figure draws every task on the same radius.
SAGE leads every one of the nineteen tasks, in domain, out of domain and on the upstream corpora that none of the compared systems has trained on. On the block averages of \Cref{tab:maeb-avg}, which pool metrics of different kinds as the MAEB leaderboards do, the five baselines lie within $8\%$ of one another and SAGE stands $10\%$ to $28\%$ above the best of them, most of all on the upstream tasks. \Cref{fig:umap} shows that geometry with no model fitted at all, projecting with UMAP \citep{umap} the descriptors of the isolated MoisesDB stems, colored by their ground-truth instrument: the stems group by
  \par
\end{minipage}\hfill
\begin{minipage}[t]{0.32\linewidth}
  \vspace{0pt}
  \centering
  \includegraphics[scale=0.95]{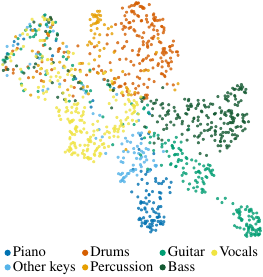}
  \setlength{\abovecaptionskip}{2pt}
  \captionof{figure}{UMAP of SAGE.}
  \label{fig:umap}
\end{minipage}

\noindent instrument, and related sources lie side by side, the bass next to the guitar, the other keys next to the piano and the percussion at the edge of the drums.

\begin{table}[!ht]
  \centering
  \footnotesize
  \setlength{\tabcolsep}{3pt}
  \renewcommand{\arraystretch}{0.9}
  \caption{Average score over the tasks of each block.}
  \label{tab:maeb-avg}
  \begin{tabular}{lcccccc}
    \toprule
    Block & \textbf{SAGE (ours)} & SAME-L & SAME-S & Stable Audio Open & CoDiCodec & Music2Latent \\
    \midrule
    FMA (6)           & $\mathbf{0.563}$ & $0.470$ & $0.468$ & $0.490$ & $0.474$ & $0.484$ \\
    MoisesDB (7)      & $\mathbf{0.544}$ & $0.493$ & $0.487$ & $0.469$ & $0.456$ & $0.479$ \\
    Upstream MAEB (6) & $\mathbf{0.622}$ & $0.473$ & $0.472$ & $0.481$ & $0.471$ & $0.487$ \\
    \bottomrule
  \end{tabular}
\end{table}

\subsection{Ablations}
\label{sec:ablations}

\begin{table}[!ht]
  \centering
  \footnotesize
  \setlength{\tabcolsep}{4pt}
  \caption{Semantic branch on FMA-medium, probes averaged over the six FMA tasks. Gate steps are optimizer steps, one generator update every three of them.}
  \label{tab:ab-semantic}
  \begin{tabular}{lrrrrrrr}
    \toprule
    & \multicolumn{2}{c}{FAD $\downarrow$} & \multicolumn{2}{c}{CLAP $\uparrow$} & & & \\
    \cmidrule(lr){2-3} \cmidrule(lr){4-5}
    Configuration & MERT & CLAP & music & audio & SI-SDR $\uparrow$ & STFT $\downarrow$ & Probes $\uparrow$ \\
    \midrule
    \multicolumn{8}{l}{\textit{Gate and strength, with the WavTokenizer discriminator}} \\
    $s_0 = 25$k, $\lambda_{\mathrm{sem}} = 1$ (adopted) & $0.240$ & $0.051$ & $0.881$ & $0.879$ & $0.76$  & $\mathbf{1.099}$ & $0.637$ \\
    $s_0 = 50$k, $\lambda_{\mathrm{sem}} = 1$           & $0.248$ & $\mathbf{0.046}$ & $\mathbf{0.889}$ & $\mathbf{0.885}$ & $0.35$  & $1.117$ & $0.629$ \\
    $s_0 = 100$k, $\lambda_{\mathrm{sem}} = 1$          & $\mathbf{0.218}$ & $0.051$ & $0.878$ & $0.881$ & $\mathbf{1.46}$  & $1.121$ & $0.614$ \\
    $s_0 = 25$k, $\lambda_{\mathrm{sem}} = 0.5$         & $0.250$ & $0.052$ & $0.885$ & $0.882$ & $1.01$  & $1.102$ & $0.631$ \\
    $s_0 = 25$k, $\lambda_{\mathrm{sem}} = 2$           & $0.382$ & $0.061$ & $0.872$ & $0.870$ & $-0.06$ & $1.134$ & $\mathbf{0.641}$ \\
    \midrule
    \multicolumn{8}{l}{\textit{Teacher, without the adversarial branch}} \\
    none           & $\mathbf{1.453}$ & $\mathbf{0.41}$ & $\mathbf{0.645}$ & $\mathbf{0.624}$ & $\mathbf{2.47}$ & $1.083$ & $0.495$ \\
    MERT           & $1.490$ & $0.44$ & $0.625$ & $0.607$ & $2.28$ & $\mathbf{1.052}$ & $0.521$ \\
    CLAP (adopted) & $1.746$ & $0.48$ & $0.613$ & $0.575$ & $1.60$ & $1.071$ & $\mathbf{0.642}$ \\
    \bottomrule
  \end{tabular}
\end{table}

\Cref{tab:ab-semantic} sweeps the two knobs of \Cref{sec:distillation} on FMA-medium, with a reduced backbone that predates the SwiGLU and XSA variants of the final one, so what transfers is the ordering inside a study and not the absolute values. The sweep traces the trade the method rests on: doubling the weight buys the best probing average of the sweep and pays for it with a FAD-MERT $60\%$ worse, opening the gate four times later reverses both, and the configuration we adopt sits at the knee. The lower block first removes the teacher and then changes it, everything else held fixed. Distillation costs fidelity on every column but the STFT distance, and what it returns is probing structure: little of it with the audio-only teacher, a great deal with the language-audio one. The CLAP-based metrics move against the model that distills CLAP on both knobs: the strongest weight gives the worst CLAP cosine and the worst FAD-CLAP of the sweep, and in the lower block the CLAP-distilled arm scores below the undistilled one on both CLAP similarities, the music one being the teacher itself. A metric inflated by its own training signal would move the other way. Removing the teacher drops the probing average from $0.642$ to $0.495$, and with it most of the margin \Cref{sec:probing} reports over the compared systems.

Appendix~\ref{app:ablations} carries two further studies.

\section{Related Work}
\label{sec:related-work}

\paragraph{Audio autoencoders.} Recent audio autoencoders either keep a continuous, typically variational, latent or discretize it through residual vector quantization \citep{soundstream, encodec, dac}. Neural codecs target the best fidelity at a given bitrate and quantize to reach bitrates a continuous latent cannot; the autoencoders behind latent diffusion stay continuous, lightly regularized towards a Gaussian prior, and trade some fidelity for a latent the second stage can model more easily. The consistency autoencoders Music2Latent and CoDiCodec decode in a single step with a consistency model \citep{consistencymodels}, trading sample-exactness for plausibility.

Audio autoencoders also differ in their input representation. Waveform-domain models such as Stable Audio Open and SAME learn phase-aware filtering from samples; models on magnitude or mel spectrograms, as in AudioLDM \citep{audioldm}, discard phase and need a separately trained vocoder; models on the complex STFT, Music2Latent, CoDiCodec and SAGE among them, recover the waveform with a single inverse transform.

\paragraph{Semantic alignment.}
\label{sec:semantic-alignment}
Most of the systems above optimize the latent for reconstruction alone, whereas recent work shapes it on purpose, arguing that the encoder is the first stage of a generative system and the geometry it leaves behind is what the second stage has to model. In audio, the idea has been pursued from two directions. SAME shapes its latent with linear regressors onto hand-crafted chroma and stereo-image targets and a contrastive critic trained jointly over latent, audio, and text. SALAD-VAE~\citep{salad} distills CLAP embeddings into a clip-level latent descriptor.

\paragraph{Music generative models.} Text-to-music systems generate in the latent of a pretrained autoencoder, and their quality is bounded by it. Discrete models such as MusicLM \citep{musiccaps} and MusicGen \citep{musicgen} run an autoregressive transformer over codec tokens; continuous ones such as AudioLDM and the Stable Audio family \citep{sa3} run latent diffusion over a VAE bottleneck. In both cases the generator must learn from the latent alone which directions correspond to instrument, genre or timbre, a task the image-domain evidence of the Introduction suggests an organized latent eases. Matching the budget and channel count of Stable Audio Open, SAGE is a drop-in first stage for diffusion architectures designed at that rate.

\section{Conclusions}

We presented SAGE, a compact variational autoencoder that shapes its latent by distilling a pretrained audio-text model on a delayed schedule, so that reconstruction settles before the semantic gradient is admitted. At $105$M parameters, SAGE runs about three times faster than Stable Audio Open, behind only the distilled SAME-S, and matches the listening-test quality of SAME-L, a model eight times larger and more than twice as slow, surpassing both on the perceptual and distributional metrics of reconstruction; it also leads all nineteen probing tasks, in domain and out of domain.

\section*{AI Use Statement}
Large language models were used extensively to write and refactor the training code. All code was reviewed and run by the authors, and the method, experiments, and analysis are the authors' own; the authors take full responsibility for the content.

\section*{Ethics Statement}
SAGE was trained solely on publicly available data, and the commercial music of the listening test was used only for evaluation. The listening test was conducted with volunteer participants who gave informed consent; no personal or sensitive data were collected, and responses were analyzed anonymously. As a first stage for generative models, SAGE contributes to making audio and music generation better and cheaper. That progress may displace the work of composers, musicians and audio engineers, and it raises questions of attribution, consent and misuse. We believe these models should be developed alongside tools that support audio professionals rather than replace them.

\section*{Reproducibility Statement}
The code, weights and evaluation harness of SAGE are available at the repository given in the Introduction (Section~\ref{sec:intro}). Since all training data are publicly available and the full configuration of both training phases is given in Appendix~\ref{app:hparams}, the released code is sufficient to reproduce our results. The random seeds used for the data splits are included in the configurations and, even without those, the code includes the procedure to regenerate equivalent splits, so any remaining variation should come from the partitioning alone.

\section*{Acknowledgements}
This work is supported through the MUR FIS2 grant n.~FIS-2023-00942
``NEXUS'' (cup B53C25001030001), and the Sapienza Seed of ERC grant
``MINT.AI'' (cup B83C25001040001).

\bibliography{sage}
\bibliographystyle{sage_arxiv}

\appendix

\clearpage
\section{Hyperparameters}
\label{app:hparams}

Table~\ref{tab:hparams} lists the complete configuration of the model reported in the paper,
together with the two training phases of Section~\ref{sec:setup}.

\begin{table}[!ht]
  \centering
  \footnotesize
  \caption{Full configuration of SAGE and of the two training phases.}
  \label{tab:hparams}
  \begin{tabular}{ll}
    \toprule
    \multicolumn{2}{l}{\textit{Input representation}} \\
    Sample rate, channels            & $44.1$\,kHz, stereo \\
    STFT window / hop                & $2048$ (Hann) / $512$ \\
    Frequency bins, frames           & $F = 1024$ (Nyquist dropped), $T = 128$ \\
    Training segment                 & $65{,}024$ samples per channel ($\approx 1.5$\,s) \\
    Power-law compression $\alpha$, scale $\beta$ & $0.65$, $0.35$ \\
    \midrule
    \multicolumn{2}{l}{\textit{Architecture}} \\
    Patch size, embedding width      & $64 \times 1$, $C = 256$ \\
    Stage depths, attention heads    & $(2, 6, 2)$, $(8, 16, 32)$ \\
    Attention window                 & $4 \times 32$ \\
    Feed-forward                     & SwiGLU, hidden ratio $3.0$ \\
    Attention variant, normalization & exclusive self-attention, post-residual \\
    Latent                           & $C_z = 16$, grid $4 \times 32$ \\
    Compression rate                 & $130{,}048$ waveform samples $\to 2{,}048$ latent scalars ($63.5\times$) \\
    Stochastic depth                 & $0.1$ \\
    Post-net (decoder only)          & $2$ convolutions, $7 \times 3$ kernels, $64$ channels \\
    Parameters                       & $104{,}629{,}380$ \\
    \midrule
    \multicolumn{2}{l}{\textit{Stereo terms}} \\
    Sum-and-difference MR-STFT       & $M$, $S$, $L$, $R$; FFT $2048$ to $64$, hop FFT$/4$, A-weighted \\
    Discriminator inputs             & $L$, $R$, $M$, $S$ folded into the batch \\
    \midrule
    \multicolumn{2}{l}{\textit{Optimization, phase 1}} \\
    Hardware, global batch           & $16$ A100 ($4 \times 4$), $128$ segments \\
    Epochs, batches                  & $500$, $956$ per epoch ($\approx 478$k batches in total) \\
    Generator updates                & $\approx 239$k, one every two batches, alternating with the discriminator \\
    Generator optimizer              & AdamW, lr $10^{-3}$, $\beta = (0.9, 0.98)$, wd $10^{-4}$ (matrices only) \\
    Schedule                         & inverse square root, $\gamma = 2 \times 10^{5}$, power $1/2$ \\
    Warm-up                          & exponential, $90\%$ of peak at $\approx 2{,}400$ steps \\
    Discriminator optimizer          & AdamW, lr $2 \times 10^{-4}$, $\beta = (0.8, 0.99)$, wd $10^{-3}$ \\
    Projection head optimizer        & AdamW, lr $10^{-5}$, constant \\
    Semantic gate $s_0$              & $\approx 8.3$k generator updates ($25{,}000$ steps over all three optimizers) \\
    Loss weights, order of \eqref{eq:total} & $(1, 0.5, 1, 10^{-4}, 1, 0.1, 0.2)$ \\
    Gradient clipping, precision     & norm $20$, TF32 \\
    EMA decay, seed                  & $0.9998$, $94$ \\
    \midrule
    \multicolumn{2}{l}{\textit{Optimization, phase 2 (differences only)}} \\
    Epochs                           & $992$ \\
    Generator learning rate          & $10^{-4}$, schedule and warm-up restarted \\
    Encoder, discriminator           & frozen, re-initialized from scratch \\
    Loss weights                     & $\lambda_{\mathrm{KL}} = \lambda_{\mathrm{sem}} = 0$, $\lambda_{\mathrm{SD}} = 1$\\
    \bottomrule
  \end{tabular}
\end{table}
\section{Stereo Image}
\label{app:stereo}

The objective of Section~\ref{sec:curriculum} carries $\mathcal{L}_{\mathrm{SD}}$ for the stereo
image alone, and Table~\ref{tab:ab-stereo} measures what the term does. The four quantities are
computed on the MoisesDB mixtures against their references, on the multi-corpus pretraining of
Section~\ref{sec:setup} rather than on the reduced setup of Section~\ref{sec:ablations}. The level
of the side, in decibels relative to the side of the reference, is zero when the image is as wide as
it should be and negative when the reconstruction is narrower; $d_{\mathrm{width}}$ is a distance
between the width envelopes of reconstruction and reference; and the last two columns are the
scale-invariant signal-to-distortion ratios of the side and of the mid, which separate the level of
the side from its content.

Without the term the side comes out about $12$\,dB too quiet, and the deficit is the same after
three hundred and after fifteen hundred epochs, which rules out an effect of under-training. With
it the level is essentially matched, the width distance falls by a third and the side gains more
than nine decibels of content, while the mid is left where it was.

\begin{table}[!ht]
  \centering
  \small
  \caption{Stereo image on the MoisesDB mixtures, with and without $\mathcal{L}_{\mathrm{SD}}$ in pretraining.}
  \label{tab:ab-stereo}
  \begin{tabular}{lrrrr}
    \toprule
    Pre-training & Side level (dB) & $d_{\mathrm{width}}$ $\downarrow$ & SI-SDR side $\uparrow$ & SI-SDR mid $\uparrow$ \\
    \midrule
    without $\mathcal{L}_{\mathrm{SD}}$, epoch $319$   & $-12.40$ & $0.092$ & $-14.81$ & $4.87$ \\
    without $\mathcal{L}_{\mathrm{SD}}$, epoch $1499$  & $-11.99$ & $0.091$ & $-14.32$ & $4.87$ \\
    with $\mathcal{L}_{\mathrm{SD}}$, epoch $499$      & $-0.20$  & $0.060$ & $-5.09$  & $4.84$ \\
    \bottomrule
  \end{tabular}
\end{table}

\section{Inference Timing and Numerical Precision}
\label{app:timing}

\paragraph{Protocol.} All models run on one NVIDIA RTX PRO 6000 Blackwell GPU (PyTorch 2.7.1, CUDA 12.8) through their authors' implementations and public weights, at batch size one, from CPU audio to CPU audio (transfers, STFTs, encoding, sampling and decoding), with no external chunking. After three warm-up calls each clip is timed three times; a run reports the median over $100$ clips of the per-clip median. Compiled settings wrap every network block in \texttt{torch.compile} (default mode, static shapes), and a compilation error fails the setting rather than falling back to eager. Each value is the median over $2$--$4$ independent runs, one fresh process per model and clip set in random order; a run is discarded and repeated if another process used the GPU or our CPU cores, judged from a GPU and CPU load log and never from the timings. One setting varies by $1.2\%$ between runs (95\% CI $0.8$--$1.4\%$), which the intervals of \Cref{tab:ratios} include.

\begin{table}[!ht]
  \centering
  \footnotesize
  \setlength{\tabcolsep}{1.8pt}
  \caption{Time relative to SAGE at the same precision, each model at its faster of eager and compiled, with 95\% intervals from clip sampling and the between-run spread. Below one: faster than SAGE.}
  \label{tab:ratios}
  \begin{tabular}{@{}lrrrrrr@{}}
    \toprule
    & \multicolumn{3}{c}{MoisesDB $10$\,s} & \multicolumn{3}{c}{FMA $30$\,s} \\
    \cmidrule(lr){2-4} \cmidrule(lr){5-7}
    Model & Pkg.\ defaults & TF32 & BF16 & Pkg.\ defaults & TF32 & BF16 \\
    \midrule
    SAME-L & $\times3.75$ {\tiny[3.65,3.85]} & $\times2.12$ {\tiny[2.07,2.18]} & $\times1.67$ {\tiny[1.63,1.72]} & $\times4.19$ {\tiny[4.08,4.31]} & $\times2.56$ {\tiny[2.49,2.63]} & $\times2.03$ {\tiny[1.97,2.08]} \\
    SAME-S & $\times0.60$ {\tiny[0.58,0.61]} & $\times0.47$ {\tiny[0.46,0.49]} & $\times0.40$ {\tiny[0.39,0.42]} & $\times0.60$ {\tiny[0.58,0.61]} & $\times0.48$ {\tiny[0.47,0.49]} & $\times0.42$ {\tiny[0.41,0.44]} \\
    Stable Audio Open & $\times1.47$ {\tiny[1.43,1.51]} & $\times2.78$ {\tiny[2.70,2.85]} & $\times2.02$ {\tiny[1.97,2.08]} & $\times1.73$ {\tiny[1.68,1.78]} & $\times3.28$ {\tiny[3.18,3.37]} & $\times3.16$ {\tiny[3.07,3.25]} \\
    \bottomrule
  \end{tabular}
\end{table}

\paragraph{Numerical precision.} \emph{Package defaults} leaves each model as its package configures it: FP32 matrix multiplications for SAGE and SAME, TF32 convolutions for Stable Audio Open, and the internal FP16 autocast of CoDiCodec and Music2Latent, kept in every setting. \emph{TF32} enables TF32 matrix multiplications and convolutions; \emph{BF16} autocasts the network blocks with FP32 weights, keeping inputs, outputs and STFTs in FP32. Each setting is compared with package defaults in eager mode on the same clips and latent samples (SI-SDR, SDR, mel and MR-STFT distances, paired 95\% bootstrap intervals): $\dagger$ marks a loss beyond both the latent-resampling noise and $1\%$, $*$ a measurable loss between $0.1\%$ and $1\%$. TF32 changes no model by more than $0.1\%$ except SAME-L ($-0.1\%$ SI-SDR); BF16 makes SAME-L, SAME-S, Stable Audio Open and CoDiCodec worse, while SAGE loses at most $0.2\%$. SAME runs flex attention without flash-attn, which costs SAME-L $7$--$9\%$ when compiled with package defaults or BF16, and under $1\%$ with TF32.

\begin{table}[!ht]
  \centering
  \footnotesize
  \setlength{\tabcolsep}{3pt}
  \caption{Inference time per clip in ms for every precision, median over $2$--$4$ runs. $\dagger$: quality worse than package defaults in eager mode; $*$: loss between $0.1\%$ and $1\%$.}
  \label{tab:precision}
  \begin{tabular}{@{}lrrrrrr@{}}
    \toprule
    & \multicolumn{2}{c}{Package defaults} & \multicolumn{2}{c}{TF32} & \multicolumn{2}{c}{BF16} \\
    \cmidrule(lr){2-3} \cmidrule(lr){4-5} \cmidrule(lr){6-7}
    Model & Eager & Comp. & Eager & Comp. & Eager & Comp. \\
    \midrule
    \multicolumn{7}{@{}l}{\emph{MoisesDB mixtures, $10$\,s}} \\
    \textbf{SAGE (ours)} & $34.2$ & $23.3$ & $20.3$ & $12.2$ & $23.6$ & $12.5$ \\
    SAME-L & $123.4$ & $87.4$ & $33.4$ & $25.9^{*}$ & $27.1^{\dagger}$ & $21.0^{\dagger}$ \\
    SAME-S & $21.1$ & $13.9$ & $8.54$ & $5.77$ & $9.89^{\dagger}$ & $5.07^{*}$ \\
    Stable Audio Open & $75.1$ & $34.2$ & $74.9$ & $33.9$ & $67.6^{\dagger}$ & $25.4^{\dagger}$ \\
    CoDiCodec & $252.0$ & $148.1$ & $251.7$ & $148.2$ & $295.4^{*}$ & $184.9^{*}$ \\
    Music2Latent & $298.4$ & $136.8$ & $286.9$ & $137.4$ & $314.0^{*}$ & $153.0$ \\
    \midrule
    \multicolumn{7}{@{}l}{\emph{FMA test, $30$\,s}} \\
    \textbf{SAGE (ours)} & $99.9$ & $57.7$ & $42.0$ & $30.6$ & $37.6^{*}$ & $25.5^{*}$ \\
    SAME-L & $422.7$ & $241.8$ & $102.0^{*}$ & $78.3^{*}$ & $79.0^{\dagger}$ & $51.8^{\dagger}$ \\
    SAME-S & $54.6$ & $34.5$ & $20.8$ & $14.7$ & $17.8^{\dagger}$ & $10.8^{\dagger}$ \\
    Stable Audio Open & $238.3$ & $100.0$ & $239.4$ & $100.1$ & $223.7^{\dagger}$ & $80.6^{\dagger}$ \\
    CoDiCodec & $799.6$ & $413.8$ & $796.8$ & $413.8$ & $906.9^{\dagger}$ & $521.1^{\dagger}$ \\
    Music2Latent & $993.6$ & $473.8$ & $990.3$ & $405.3$ & $1022.0$ & $509.6$ \\
    \bottomrule
  \end{tabular}
\end{table}

SALAD-VAE, whose objective \Cref{sec:distillation} adapts, is excluded from the comparison because it has released neither weights nor code. It targets general audio and reports fidelity on speech rather than music.

\section{Semantic Probing by Task}
\label{app:maeb}

Table~\ref{tab:maeb-full} gives the absolute scores behind Figure~\ref{fig:maeb-radar} and
Table~\ref{tab:maeb-avg}, each under the metric MAEB defines for its task type and with the protocol of
Section~\ref{sec:protocol}; retrieval is audio-to-audio throughout.

\begin{table}[!ht]
  \centering
  \scriptsize
  \setlength{\tabcolsep}{1.5pt}
  \caption{Semantic probing on the nineteen tasks, best per row among the six autoencoders in bold, block averages in the last row of each block. $^{*}$SAME-L and SAME-S are probed at their native $256$ dimensions, four times the width of every other descriptor. $^{\dagger}$CLAP-oracle, underlined throughout, probes the frozen $512$-dimensional teacher of \Cref{sec:distillation} under the same protocol; it cannot be inverted back to audio and is a reference, not a seventh autoencoder, so it is excluded from the bolding.}
  \label{tab:maeb-full}
  \begin{tabular}{llrrrrrrr}
    \toprule
    Task & Metric & \textbf{SAGE} & SAME-L$^{*}$ & SAME-S$^{*}$ & \shortstack[r]{Stable\\Audio Open} & CoDiCodec & \shortstack[r]{Music2-\\Latent} & \shortstack[r]{CLAP-\\oracle$^{\dagger}$} \\
    \midrule
    \multicolumn{9}{l}{\textit{FMA test split, in domain}} \\
    Genre classification & accuracy & $\mathbf{0.447}$ & $0.285$ & $0.279$ & $0.287$ & $0.305$ & $0.307$ & $\underline{0.529}$ \\
    Genre clustering & V-measure & $\mathbf{0.215}$ & $0.146$ & $0.145$ & $0.171$ & $0.168$ & $0.185$ & $\underline{0.385}$ \\
    Artist clustering & V-measure & $\mathbf{0.706}$ & $0.634$ & $0.634$ & $0.655$ & $0.617$ & $0.632$ & $\underline{0.719}$ \\
    Artist retrieval & nDCG@10 & $\mathbf{0.583}$ & $0.442$ & $0.438$ & $0.486$ & $0.448$ & $0.435$ & $\underline{0.587}$ \\
    Genre reranking & MAP@1000 & $\mathbf{0.535}$ & $0.491$ & $0.491$ & $0.493$ & $0.493$ & $0.513$ & $\underline{0.626}$ \\
    Artist pair classification & max AP & $\mathbf{0.893}$ & $0.824$ & $0.822$ & $0.849$ & $0.812$ & $0.833$ & $\underline{0.935}$ \\
    \textit{Average} &  & $\mathbf{0.563}$ & $0.470$ & $0.468$ & $0.490$ & $0.474$ & $0.484$ & $\underline{0.630}$ \\
    \midrule
    \multicolumn{9}{l}{\textit{MoisesDB, out of domain}} \\
    Genre classification & accuracy & $\mathbf{0.533}$ & $0.453$ & $0.444$ & $0.409$ & $0.400$ & $0.427$ & $\underline{0.698}$ \\
    Genre clustering & V-measure & $\mathbf{0.193}$ & $0.157$ & $0.141$ & $0.155$ & $0.112$ & $0.132$ & $\underline{0.351}$ \\
    Artist clustering & V-measure & $\mathbf{0.532}$ & $0.466$ & $0.475$ & $0.479$ & $0.467$ & $0.472$ & $\underline{0.612}$ \\
    Artist retrieval & nDCG@10 & $\mathbf{0.574}$ & $0.524$ & $0.517$ & $0.517$ & $0.491$ & $0.518$ & $\underline{0.693}$ \\
    Genre reranking & MAP@1000 & $\mathbf{0.607}$ & $0.588$ & $0.580$ & $0.561$ & $0.557$ & $0.573$ & $\underline{0.682}$ \\
    Artist pair classification & max AP & $\mathbf{0.744}$ & $0.697$ & $0.695$ & $0.679$ & $0.648$ & $0.680$ & $\underline{0.826}$ \\
    Instrument classification (stems) & accuracy & $\mathbf{0.624}$ & $0.563$ & $0.560$ & $0.480$ & $0.516$ & $0.555$ & $\underline{0.695}$ \\
    \textit{Average} &  & $\mathbf{0.544}$ & $0.493$ & $0.487$ & $0.469$ & $0.456$ & $0.479$ & $\underline{0.651}$ \\
    \midrule
    \multicolumn{9}{l}{\textit{Upstream MAEB tasks, on their original datasets}} \\
    GTZAN genre classification & accuracy & $\mathbf{0.774}$ & $0.531$ & $0.530$ & $0.500$ & $0.529$ & $0.494$ & $\underline{0.840}$ \\
    GTZAN genre clustering & V-measure & $\mathbf{0.471}$ & $0.272$ & $0.269$ & $0.311$ & $0.337$ & $0.336$ & $\underline{0.679}$ \\
    Music Genre clustering & V-measure & $\mathbf{0.344}$ & $0.126$ & $0.125$ & $0.139$ & $0.182$ & $0.144$ & $\underline{0.450}$ \\
    GTZAN genre reranking & MAP@1000 & $\mathbf{0.722}$ & $0.555$ & $0.554$ & $0.557$ & $0.588$ & $0.573$ & $\underline{0.816}$ \\
    NSynth instrument source & accuracy & $\mathbf{0.453}$ & $0.396$ & $0.396$ & $0.419$ & $0.432$ & $0.406$ & $\underline{0.595}$ \\
    Jam-ALT artist retrieval$^{\ddagger}$ & nDCG@10 & $\mathbf{0.9672}$ & $0.9601$ & $0.9592$ & $0.9611$ & $0.7603$ & $0.9668$ & $\underline{0.923}$ \\
    \textit{Average} &  & $\mathbf{0.622}$ & $0.473$ & $0.472$ & $0.481$ & $0.471$ & $0.487$ & $\underline{0.717}$ \\
    \bottomrule
  \end{tabular}
  \par\smallskip
  $^{\ddagger}$Autoencoders at four decimals, since SAGE and Music2Latent coincide at three.
\end{table}

Averaged over the nineteen tasks, SAGE retains $84\%$ of the CLAP-oracle score. The gap concentrates in genre clustering, whose tasks give three of the suite's four lowest figures, from $55\%$ to $69\%$, and $76\%$ on Music Genre, the same level as MoisesDB genre classification and NSynth; artist clustering, equally training-free, retains $87\%$ to $98\%$, so the drop tracks genre rather than the absence of supervision. Every other task retains above $82\%$, and on Jam-ALT artist retrieval SAGE exceeds the oracle outright.

\section{Additional Ablations}
\label{app:ablations}

The two studies below isolate one factor at a time on FMA-medium, with a reduced backbone and, where indicated, no adversarial branch; the winner of each is what Section~\ref{sec:setup} adopts. As in the semantic study of Section~\ref{sec:ablations}, the runs predate the SwiGLU and XSA variants of the final backbone, so what transfers is the ordering inside a study, not the absolute values.

\paragraph{Adversarial branch.} Any discriminator moves the distributional and perceptual metrics a long way, the WavTokenizer ensemble cutting FAD-MERT sevenfold at a cost in SI-SDR, and the objective matters as much as the architecture: on the same EnCodec critic the relativistic formulation reaches $0.387$ where the hinge loss stops at $1.048$ and drives SI-SDR negative. The lower block of Table~\ref{tab:ab-disc} moves the same critics between the two phases, and the best moment to introduce one depends on which one it is. The WavTokenizer ensemble is best present from the first step, $0.208$ against $0.419$ when it arrives only at fine-tuning, whereas the EnCodec-style critic is better introduced late, $0.548$ against $0.959$; resuming its weights into the second phase rather than re-initializing them collapses every column.

\begin{table}[!ht]
  \centering
  \footnotesize
  \setlength{\tabcolsep}{4pt}
  \caption{Discriminator type and placement on FMA-medium; only the first two rows of the lower block distill a teacher.}
  \label{tab:ab-disc}
  \begin{tabular}{llrrrrr}
    \toprule
     &  & \multicolumn{2}{c}{FAD $\downarrow$} & \multicolumn{2}{c}{CLAP $\uparrow$} &  \\
    \cmidrule(lr){3-4} \cmidrule(lr){5-6}
    \multicolumn{2}{l}{Discriminator} & MERT & CLAP & music & audio & SI-SDR $\uparrow$ \\
    \midrule
    \multicolumn{7}{l}{\textit{Type, present from the first step}} \\
    \multicolumn{2}{l}{WavTokenizer-RpGAN} & $\mathbf{0.203}$ & $\mathbf{0.056}$ & $\mathbf{0.881}$ & $\mathbf{0.879}$ & $1.55$ \\
    \multicolumn{2}{l}{EnCodec-RpGAN} & $0.387$ & $0.122$ & $0.821$ & $0.804$ & $0.55$ \\
    \multicolumn{2}{l}{EnCodec-hinge} & $1.048$ & $0.265$ & $0.693$ & $0.660$ & $-2.21$ \\
    \multicolumn{2}{l}{none} & $1.453$ & $0.414$ & $0.645$ & $0.624$ & $\mathbf{2.47}$ \\
    \midrule
    \multicolumn{7}{l}{\textit{Placement across the two phases}} \\
    Pre-training & Fine-tuning &  &  &  &  &  \\
    \cmidrule(lr){1-2}
    WavTokenizer & WavTokenizer & $\mathbf{0.208}$ & $\mathbf{0.045}$ & $\mathbf{0.892}$ & $\mathbf{0.889}$ & $0.80$ \\
    none & WavTokenizer & $0.419$ & $0.064$ & $0.860$ & $0.864$ & $0.49$ \\
    none & EnCodec-hinge & $0.548$ & $0.130$ & $0.819$ & $0.813$ & $\mathbf{2.49}$ \\
    EnCodec-hinge & EnCodec-hinge, reset & $0.959$ & $0.200$ & $0.730$ & $0.701$ & $-1.98$ \\
    EnCodec-hinge & EnCodec-hinge, resumed & $2.553$ & $0.450$ & $0.473$ & $0.465$ & $-7.87$ \\
    \bottomrule
  \end{tabular}
\end{table}

\paragraph{Capacity.} Table~\ref{tab:ab-scale} grows the backbone along one axis at a time. Deepening the middle stage from six blocks to eighteen, the step Swin itself takes between its tiny and its small configuration, adds $83$M parameters and improves both Fr\'echet distances and both CLAP similarities. Widening the embedding to $C = 512$ costs four times the parameters and improves FAD-CLAP by a quarter and the two CLAP similarities, conceding a little on FAD-MERT. Neither change asks for a different architecture, and the return per parameter is larger along the depth axis; the configuration we report sits at the compact end of both curves.

\begin{table}[!ht]
  \centering
  \footnotesize
  \setlength{\tabcolsep}{4pt}
  \caption{Capacity on FMA-medium, without the adversarial branch; values compare inside a study, not across the two.}
  \label{tab:ab-scale}
  \begin{tabular}{lrrrrrr}
    \toprule
     &  & \multicolumn{2}{c}{FAD $\downarrow$} & \multicolumn{2}{c}{CLAP $\uparrow$} &  \\
    \cmidrule(lr){3-4} \cmidrule(lr){5-6}
    Configuration & Params & MERT & CLAP & music & audio & SI-SDR $\uparrow$ \\
    \midrule
    \multicolumn{7}{l}{\textit{Width, $\lambda_{\mathrm{mel}} = 0.5$}} \\
    $C = 512$, $(2, 6, 2)$ & $388$M & $1.050$ & $\mathbf{0.259}$ & $\mathbf{0.700}$ & $\mathbf{0.720}$ & $1.58$ \\
    $C = 256$, $(2, 6, 2)$ & $97$M & $\mathbf{0.963}$ & $0.347$ & $0.680$ & $0.674$ & $\mathbf{1.68}$ \\
    \addlinespace
    \multicolumn{7}{l}{\textit{Depth, $\lambda_{\mathrm{mel}} = 0.3$}} \\
    $C = 256$, $(2, 18, 2)$ & $180$M & $\mathbf{1.298}$ & $\mathbf{0.38}$ & $\mathbf{0.667}$ & $\mathbf{0.650}$ & $\mathbf{2.94}$ \\
    $C = 256$, $(2, 6, 2)$ & $97$M & $1.453$ & $0.41$ & $0.645$ & $0.624$ & $2.47$ \\
    \bottomrule
  \end{tabular}
\end{table}

\section{Training Objective}
\label{app:losses}

Let $a \in \mathbb{R}^{2 \times N}$ be a stereo segment with channels $L, R$, and let $M = L + R$, $S = L - R$; hats mark the reconstructed waveform $\hat{a}$ and the signals derived from it. Throughout, $\|\cdot\|_1$ and $\|\cdot\|_2^2$ denote means over all entries (batch, channels, frequency, time; $|\cdot|^2$ for complex entries), $\|\cdot\|_F$ the Frobenius norm over the frequency-time plane of a single signal, and $\mathbb{E}$ the mean over the batch.

\paragraph{Spectral reconstruction.} With $c(Z) = \beta\,|Z|^{\alpha} e^{i \angle Z}$, $\alpha = 0.65$, $\beta = 0.35$, let $\mathbf{X} = c\big(\mathrm{STFT}(a)\big)$ be the power-compressed complex spectrogram of \Cref{sec:backbone} (window $2048$, hop $512$, Nyquist bin dropped), whose real and imaginary parts form the encoder input. The decoder predicts $\hat{\mathbf{X}} = D(z)$ in the same compressed domain, the waveform is recovered as $\hat{a} = \mathrm{iSTFT}\big(c^{-1}(\hat{\mathbf{X}})\big)$ with $c^{-1}(Z) = (|Z|/\beta)^{1/\alpha} e^{i \angle Z}$, and
\begin{equation}
  \mathcal{L}_{\mathrm{STFT}} = \big\| \hat{\mathbf{X}} - \mathbf{X} \big\|_2^2 ,
\end{equation}
which penalizes magnitude and phase jointly.

\paragraph{Mel reconstruction.} For windows $w \in \{2048, 1024, 512\}$, $\mathrm{Mel}_w$ projects the magnitude STFT (Hann window $w$, hop $w/4$) onto $128$ mel bands from $0$\,Hz to Nyquist, and with $\epsilon = 10^{-5}$
\begin{equation}
  \mathcal{L}_{\mathrm{mel}} = \sum_{w} \Big( \big\| \mathrm{Mel}_w(a) - \mathrm{Mel}_w(\hat{a}) \big\|_1 + 2\,\big\| \log_{10} \max\!\big(\mathrm{Mel}_w(a), \epsilon\big) - \log_{10} \max\!\big(\mathrm{Mel}_w(\hat{a}), \epsilon\big) \big\|_1 \Big) .
\end{equation}

\paragraph{Sum-and-difference loss.} Let $Y_n$ be the STFT (Hann window $n$, hop $n/4$) of the signal $y$ after A-weighting, with magnitudes floored at $10^{-4}$. Over the four signals and six FFT sizes, $n$ doubling from $64$ to $2048$,
\begin{equation}
  \mathcal{L}_{\mathrm{SD}} = \frac{1}{2} \sum_{y \in \{M, S, L, R\}} \frac{1}{6} \sum_{n} \left[ \mathbb{E}\, \frac{\big\| |\hat{Y}_n| - |Y_n| \big\|_F}{\big\| |\hat{Y}_n| \big\|_F} + \big\| \ln |Y_n| - \ln |\hat{Y}_n| \big\|_1 \right] .
\end{equation}

\paragraph{Why a squared error neglects the stereo image.} Let $e_L = \hat{\mathbf{X}}_L - \mathbf{X}_L$ and $e_R = \hat{\mathbf{X}}_R - \mathbf{X}_R$ be the errors on the two channels. For any two vectors,
\begin{equation}
  \|e_L\|_2^2 + \|e_R\|_2^2 = \tfrac{1}{2}\big(\|e_L + e_R\|_2^2 + \|e_L - e_R\|_2^2\big),
\end{equation}
so $\mathcal{L}_{\mathrm{STFT}}$ weighs the sum and the difference of the channel errors equally, and the difference, which carries the side, enters only through its squared norm, a small share of the total when the side is weak. On music, where $S$ is typically much weaker than $M$, even discarding the side barely moves the loss. The spectral-convergence term of $\mathcal{L}_{\mathrm{SD}}$ instead divides by the norm of the reconstructed magnitude of each signal, so a side shrunk towards zero inflates the term, while a near-mono reference keeps it at most about one (Appendix~\ref{app:stereo}).

\paragraph{Adversarial terms.} We use the discriminator of WavTokenizer, but keep its multi-period branch only once, since its DAC component repeats it. The result has eleven sub-discriminators $\mathcal{D}_1, \dots, \mathcal{D}_{11}$. Five are multi-period discriminators on the waveform, with periods $2, 3, 5, 7, 11$. Three are multi-resolution discriminators on the magnitude STFT, with rectangular windows of $1024$, $2048$ and $512$ samples and a hop of a quarter of the window. The last three are the multi-band discriminators of DAC on the complex STFT, with FFT sizes $2048$, $1024$, $512$ and five frequency bands each; their input has its DC offset removed and is peak-normalized. The four signals $L$, $R$, $M$, $S$ are folded into the batch, so every sub-discriminator judges each of them as a separate mono signal. Each $\mathcal{D}_b$ returns a map of logits, and the relativistic pairing compares every real segment with its own reconstruction, position by position. With $f(t) = \log(1 + e^{t})$,
\begin{align}
  \mathcal{L}_{\mathrm{disc}} &= \frac{1}{11}\sum_{b=1}^{11} \mathbb{E}\big[ f\big(\mathcal{D}_b(\hat{a}) - \mathcal{D}_b(a)\big) \big], \\
  \mathcal{L}_{\mathrm{adv}} &= \frac{1}{11}\sum_{b=1}^{11} \mathbb{E}\big[ f\big(\mathcal{D}_b(a) - \mathcal{D}_b(\hat{a})\big) \big],
\end{align}
where the mean runs over the batch and the logit positions. Feature matching compares the $K_b$ hidden feature maps $\mathcal{D}_b^{(k)}$ that sub-discriminator $b$ computes before its logits, averaging the distances first within each sub-discriminator and then across the eleven,
\begin{equation}
  \mathcal{L}_{\mathrm{fm}} = \frac{1}{11} \sum_{b=1}^{11} \frac{1}{K_b} \sum_{k=1}^{K_b} \big\| \mathcal{D}_b^{(k)}(a) - \mathcal{D}_b^{(k)}(\hat{a}) \big\|_1 .
\end{equation}

\end{document}